# The orientation of Jesuit churches in the Chiquitos missions of eastern Bolivia

## Alejandro Gangui

Universidad de Buenos Aires, Facultad de Ciencias Exactas y Naturales, Argentina.
CONICET - Universidad de Buenos Aires, Instituto de Astronomía y Física del Espacio (IAFE), Argentina.

**Abstract**

The Jesuit missions in South America were an important and unique advance in Christian evangelisation on the continent until the expulsion of the Order in 1767. Although the history and cultural aspects of these missions and their most iconic buildings have been extensively studied, the archaeoastronomy of the Guaraní peoples of the Province of Paraquaria (Province of Paraguay) has only been recently considered, with the existing studies focusing primarily on the orientation of their churches. The paper presented here, which is the first archaeoastronomical study of the Jesuit missions of Chiquitos in eastern Bolivia, is an attempt to continue and complement the previous archaeoastronomical studies of the region. The methodologies employed involved the analysis of the Jesuit churches that currently exist in this region, namely, the on-site measurements of the orientations of eight churches currently standing and the ruins of a ninth church of which only a free-standing bell tower and parts of the side walls remain preserved. The orientation measurements of a tenth church, which we were unable to visit, were determined via the use of satellite maps. The landscape surrounding these churches was then examined in detail and furthermore, a detailed cultural and historical study of the characteristics of the villages where the churches are located was carried out. Our results show that, unlike the churches of the Province of Paraquaria where meridian orientations in the north-south direction stand out, half of the studied Chiquitan Jesuit churches have shown potential canonical orientations that seem to be aligned to solar phenomena, with three exhibiting precise equinoctial orientation. In the paper we propose reasons for these orientations, including the possible relevance of illumination effects on significant internal elements within the churches – effects that were generally sought in Baroque church architecture.

## Introduction

The orientation of historic Christian churches has for many years served as a fertile field for academic study from the perspective of cultural astronomy. In recent times, this topic has received a new impetus in the literature, since deliberate orientations towards astronomical phenomena became recognised as key features of the architecture of these churches. Based on the texts of the early Christian writers and apologists, we know that the symmetrical axes of religious buildings and the apses of old churches should lie within a well-defined direction so that the priest could stand facing eastward during services (McCluskey 2015, 1704; Gangui *et al.* 2016).

However, a distinct feature of Greek temples is that the entrance is generally placed on the east side, with images of divine figures positioned on the opposite wall facing towards the rising Sun in order to offer protection against forces of death associated with the west. A long tradition of Christian churches and cathedrals display a similar type of orientation, such as Saint John Lateran or Saint Peter's in Rome. In North Africa, despite Roman dominance, one can notice numerous examples that seem to follow the same orientation pattern. For instance, in the African regions of Proconsularis and Tripolitania, one could find a number of old churches with orientations towards the west, which was a usual custom in early Christianity (Esteban *et al.* 2001, 81; Belmonte *et al.* 2006, 79). We have also found results very similar to those in our recent research in the Canary Islands, particularly in the historic city of La Laguna in the island of Tenerife (Gangui and Belmonte 2018). It should also be noted that most of the churches studied in previous literature were oriented roughly within the solar range, that is, with orientations that fall between the winter and summer solstices (for instance, the Jesuit mother church of Il Gesú in Rome which points towards the eastern horizon), with a noticeable clustering around the equinoxes and, sometimes, also around the solstices (González-García 2015, 272).

The most direct influence for the construction of churches in our study area could probably be traced back to the sixteenth century after the Council of Trent (1545-1563). Following this, Cardinal Carlos Borromeo published his *Instructiones fabricae et supellectilis ecclesiasticae* (1577), which was widely disseminated. In these instructions he indicated the direction for the main altar:

> Now, the placement of this altar must be elected at the head of the church, at the highest place where the main door is located; its back wall [of the church] points in a straight line towards the east, even though the people houses be at the back. And never orient it [the back wall] towards the solstitial east, but towards the equinoctial east […]  (Borromeo 1985 [1577]: 15, translation by author)

We know, however, that although Borromeo suggested that the church be built in such a manner so that it does not depart from ancient custom and approved tradition, many of his initial followers might have later dropped the stipulation for eastern orientation.

In this paper we focus on the Jesuit missionary churches in South America, which for almost two centuries were the most representative constructions in the process of Christian evangelisation on the continent until the Order's expulsion in 1767. The main research objective is to discern possible patterns of orientations in the studied structures and to assess whether these orientations are related to the location of the Sun or other celestial bodies when crossing the horizon, which could yield important information pertaining to their construction. Although extensive and detailed historical and cultural studies of missionary peoples and their most emblematic buildings throughout this region have been carried out, the orientation of churches in these villages had not been the subject of in-depth study until a couple of years ago (Giménez Benítez *et al.* 2018). Additionally, archaeoastronomical fieldwork that considers the urban characteristics and the writings and chronicles of the missionaries have only recently been undertaken in the Guaraní peoples' territories, which today are distributed over a large region that spreads across three countries – the northeast of Argentina, southern Brazil and Paraguay.

Intending to continue and complement the pioneering work carried out by Giménez Benítez *et al.* (2018) which provided a detailed study of Jesuit churches in the Province of Paraquaria, this paper presents an archaeoastronomical study of the orientations of Jesuit churches in nearby Chiquitos (today the Chiquitanía region). The fieldwork consisted of precise measurements of the orientations of eight churches that still exist in the region and of the ruins of a ninth church, San Juan Bautista de Taperas, of which only the monumental bell tower is still standing. Through the use of satellite maps and the reconstruction of local topography from a digital terrain model, we were able to determine the orientation data of a tenth church, Santo Corazón de Jesús, located in an area of very difficult access. Finally, all the measurements were then corroborated with the same digital terrain models.

As is well-known, Jesuit reductions (settlements for indigenous people organised and administered by the Jesuit priests) also spread out into the interior regions of the South American continent, beyond Paraguay and above the Tropic of Capricorn and were mostly located within the current limits of Bolivia (Figure 1). The missions of Moxos which were established in this area were mainly concentrated in the central north of the country in the Department of Beni (Limpias Ortiz 2007) and those of the Chiquitos occupied the northeast of Santa Cruz de la Sierra. Both groups of missions were located in the vast area currently known as eastern Bolivia (*el oriente boliviano*). However, the peoples of these different regions kept their own characteristics and particularities, due firstly to the variety of cultural origins of the local populations (the Chiquitan nations, for example, originate from the Arawak trunk, unlike the Guaraní of the Province of Paraguay) and secondly due to both the diverse geography of its sites, and aspects related to the climate.

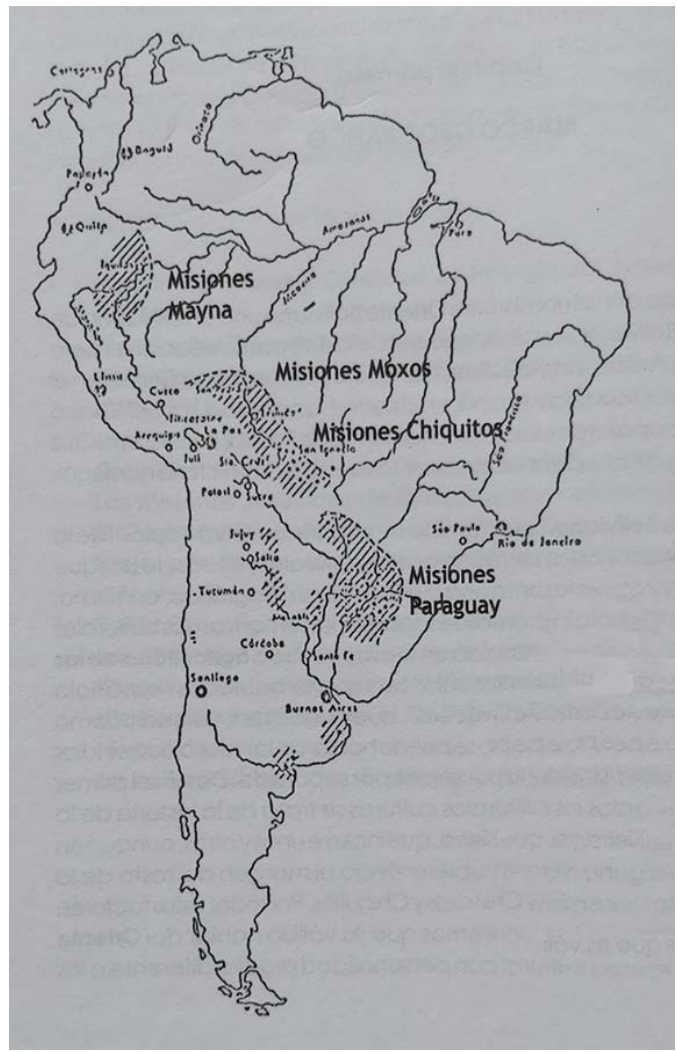

Figure 1. Map with the location of the main Jesuit missions in South America. Adapted from A. Parejas Moreno and V. Suárez Salas, *Localización de las misiones jesuíticas de Sudamérica*, 1992. In *Chiquitos. Historia de una utopía* (Santa Cruz de la Sierra: Universidad Privada de Santa Cruz de la Sierra, 1992), p. 18. To these must be added, among others, the groups of missions in Orinoco (today in Colombia and Venezuela) and at Meta in the current Colombian territory (Martini 1977, 7). (Image courtesy of A. Parejas Moreno and V. Suárez Salas).

On the other hand, the Jesuit missions of eastern Bolivia had different origins and were administered by different Jesuit provinces; those of Moxos were administered from Lima and Juli in Peru, while those of Chiquitos had their administrators in Córdoba – today in Argentinian territory – and in Tarija. It is partly due to these reasons that historians treat the two *cruceñas* reductions as separate objects of study and as interesting cases very different from those of the Jesuit peoples of the Province of Paraquaria (Parejas Moreno 1995, 253).

In the following sections we will present the historical and cultural elements that have characterised the Jesuit settlement in eastern Bolivia (Querejazu 1995). We will then review the foundation of villages, their urbanisation and the establishment of their religious buildings, especially the churches that, as we will see, represented a sort of "cathedrals" hidden in the tropical virgin forest. We will then present the data from our measurements and their analysis. In this way we will try to understand the reasons that guided the builders of the studied churches to locate and orient these buildings in the way they did. To finish, we will briefly discuss the possible relevance of internal illumination effects in the churches in determining the chosen orientations.

**The Jesuits of Chiquitos**

The Society of Jesus was created in 1540 and barely 28 years later Jesuits started carrying out their missionary activities in the Viceroyalty of Peru. A short time later they established foundations in strategic places of the Audiencia de Charcas, such as in Potosí, in La Paz and, in 1587, in Santa Cruz de la Sierra. Initially, the Jesuits concentrated their evangelising mission on the Chiriguanos of the Andean region, but towards the end of the seventeenth century a series of circumstances led them to take over the Province of Chiquitos, which is located in what is presently known as eastern Bolivia. The Chiquitanos, who were "always at war" according to Charlevoix (1913 [1757], 167), had become a serious threat to the government of the city of Santa Cruz, yet at the same time, these natives were easy prey for the slave traders from Santa Cruz, who captured and sold them in nearby regions, even in Peru. In addition, the incursions of the "flag-carriers" (*bandeirantes paulistas*), who were fortune hunters and slavers, showed the weakness of the governing body of that city which was located at the border area of Chiquitos.

Agustín de Arce y de la Concha, who was governor of Santa Cruz between 1686 and 1691, believed that the problem of the Chiquitanos could be solved through evangelisation and for this he turned to the Society of Jesus. During that time Father José de Arce, a Jesuit from La Palma in the Canary Islands, was already at the College of Tarija training himself for converting the fierce Chiriguanos to the faith of Christ. Learning of the governor's request, Father Arce immediately offered to work in Chiquitos and, with the permission of the provincial Father Gregorio Orosco who at that time was visiting Tarija, he changed his destination to first Santa Cruz and then the lands of the Chiquitanos (Hoffmann 1979, 169).

Problems began as soon as Father Arce arrived in Santa Cruz. The presence of missionaries in the city was not viewed favorably by locals, as it jeopardised the profitable business of selling young Indians as slaves. Despite some opposition from the locals and the imminence of the rainy season, which would have made his journey more difficult, Father Arce made his way towards the unmapped lands on September $2^{nd}$, 1691, accompanied by another brother of his Order and by two indigenous guides. It was not without difficulty that he arrived at the first village in Chiquitos and he did so at an opportune moment, since the Indians had been enduring the ravages caused by a plague, most likely an epidemic of smallpox, and the presence of the missionary and his help was very well received (Fernández 1895 [1726], I, Chap. IV, V.I-85).

According to Fernández, when the locals asked Father Arce to stay with them, he interpreted this as a divine sign (Fernández 1895 [1726], I, Chap. IV, V.I-87), and decided to stay in the village. Then on Saint Sylvester's Day, which coincided with the last day of the year 1691, Father Arce established the first Chiquitos reduction, which would later be dedicated to San Francisco Javier (Francis Xavier) (Parejas Moreno 1995, 273). With greater or lesser difficulties, over the years and through the intervention of many Jesuit priests, the Chiquitos territory was in due course populated by a multitude of reductions for the indigenous communities (Figure 2). Under a rigid and disciplined organisation, characteristic of the Society of Jesus, the Indians in these villages found security for their families and material well-being, but the price was absolute dedication to Roman Catholicism, complete acculturation and the total annulment of their cultural identity.

For several different reasons, such as the geographical location in adverse territories, the lack of essential natural resources like water, and the continuous attacks of the *paulistas* or of other partialities of hostile Indians, the mission of San Francisco Javier and most of those that populated the Chiquitanía, were forced to move more than once throughout the decades, even after the expulsion of the Jesuits in 1767. In many cases the dates of the foundation of the missions and even their original sites are uncertain but, excluding San Juan Bautista de Taperas, the nine missions that still stand today are "living villages" that preserve many of the traditions of centuries ago and, especially, the religious fervor of the Jesuit era.

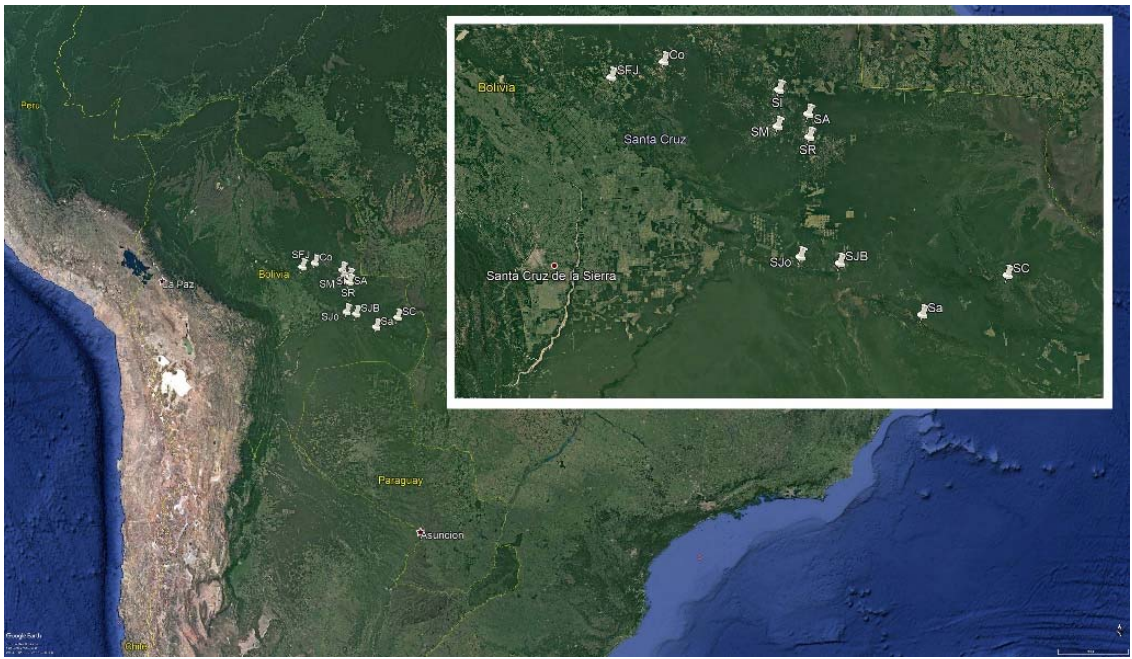

Figure 2. The ten Jesuit missions of the Province of Chiquitos, in eastern Bolivia. On the map we can see the location of the churches of San Francisco Javier (SFJ), Nuestra Señora de la Inmaculada Concepción (Co), San Ignacio (SI), San Miguel (SM), Santa Ana (SA), San Rafael (SR), San José (SJo), San Juan Bautista (SJB), Santiago (Sa) and Santo Corazón (SC). Image by the author (Google Earth Pro).

**The location sites and the urbanisation of villages**

Shortly after the occupation of the Chiquitan territory began, the Jesuits learned from the local Indians about the most suitable sites for the installation of the new reductions. To establish their villages the missionaries chose "some flat places at the extremities of the hills, as isthmus, so that on three parts they were separated by valleys from the other hills and thus had an open horizon that would provide them with a healthy open air" (Grondona 1942, 92). In many instances the location of missions took advantage of the local topography and the depths of the nearby valleys, controlling in this way the levels and gradients of a network of small streams in shallow areas (Suárez Salas 1995, 418). This made it possible to build some artificial lagoons that supplied the reductions with uninterrupted water for family consumption and agricultural production.

The organisation of the urban space in the missionary villages had its centre in the main square (*la plaza*) and next to it the temple, the college-residence and the cemetery were erected. These constructions, together with the bell tower, formed the essential religious nucleus in the configuration of the space (Gutiérrez da Costa and Gutiérrez Viñuales 1995, 338). To install the reductions, Father Knogler pointed out:

> The forest is cut down in a wide enough area and the woods and brush are burned; thus a quadrangle is drawn with a large square in the middle, three or four hundred meters long and the same in width; around the square the houses of the indians rise [...] Three sides of the square are occupied by these houses, the fourth side is reserved for the church, the cemetery, and the college where the missionaries live and where the workshops and the school are located. (Knogler 1768 [1979], chap II, 147)

The monumental character of many of the missions was notorious. Even decades after the Jesuits' estrangement from Chiquitos, ordered by decree of Charles III of Spain, travellers were surprised by the size and quality of the main squares. Alcides D'Orbigny, who visited San José de Chiquitos in 1831, said that its plaza "is enormous, decorated in the center with a stone cross surrounded by palm trees" (D'Orbigny 1945, tome III, 1180) (Figure 3).

The urban planning of these Jesuit missions expresses an alternative model to the project of the usual Hispanic city in America. In the Instructions issued in 1609 by the first Provincial Father Diego de Torres, it was indicated that the villages to be established should allocate a block for every four Indian families with a piece of land for each and that each house should have a small kitchen garden, in clear compliance with Ordinance 127 promulgated by Philip II of Spain in 1573 which stated: "The lots should be distributed to the inhabitants in parts, continuing from those corresponding to the main square" (Ordenanzas 1973, No. 127). However, in contrast to the described plan and for the first time in the New World, the lots allocated to Indian inhabitants in the Jesuit missionary villages of America were not divided into blocks. Instead, they were assigned pavilions or strips of houses that were distributed surrounding the central square (Suárez Salas 1995, 410). Likewise, regarding the property of the Indians, the distribution of land was different from the traditional one and a distinct system of common lands and collective housing was also implemented.

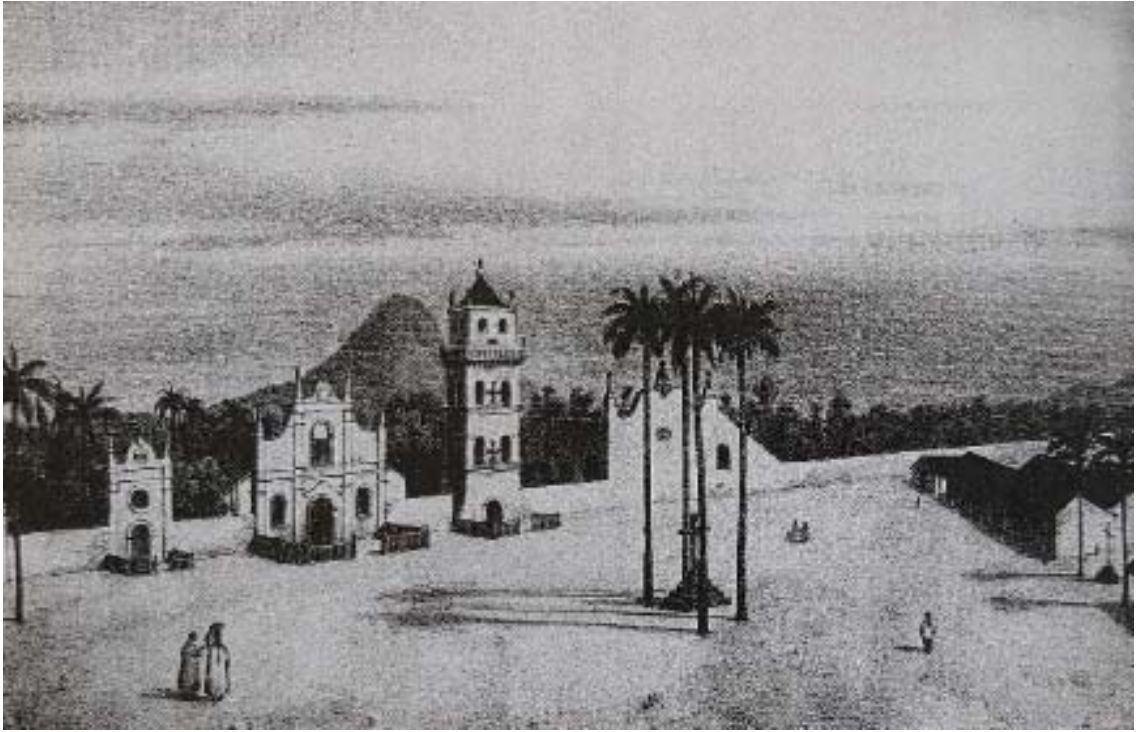

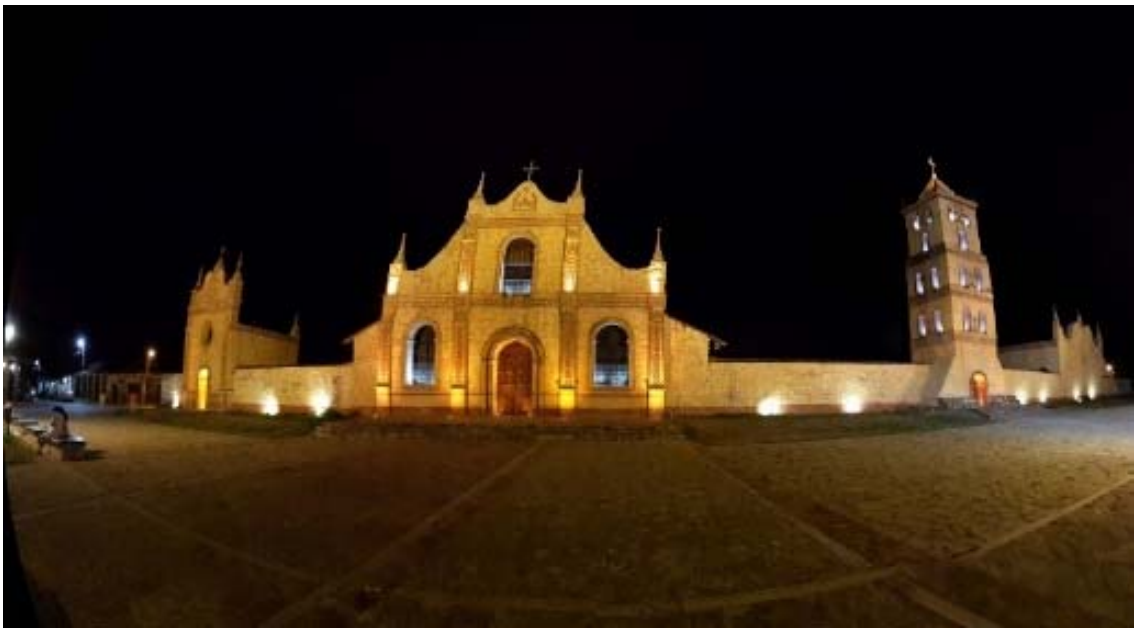

Figure 3. Top: view of the plaza of San José de Chiquitos in a drawing by D'Orbigny from 1831. The Turubó hill stands out in the background (adapted from R. Gutiérrez da Costa and R. Gutiérrez Viñuales, *Vista de la plaza de San José de Chiquitos. Alcides d'Orbigny*, 1995. In "Territorio, Urbanismo y Arquitectura en Moxos y Chiquitos", in *Las Misiones Jesuíticas de Chiquitos*, edited by P. Querejazu (La Paz: Fundación Banco Hipotecario Nacional (BHN), 1995), p. 362. Bottom: the front of the religious complex of San José today taken at night. San José stands out from the other heritage churches for its solid construction in masonry and brick and for its typically Baroque Europeanising profile. Photograph by the author. (The first image is courtesy of P. Querejazu Leyton).

The squares in Chiquitos often had rectangular proportions, were surrounded by perimeter roads and other streets that converged at the centre of the square from three different directions. The Ordinances dictated that "Four main streets emerge from the plaza, one on each side of the plaza and two streets for each corner" (Ordenanzas 1973, No. 114). However, this rule was modified even in the earliest of the established missionary villages - in particular, the street on the side of the religious complex was eliminated. From a cultural point of view, Chiquitan urbanism can be understood as manifesting in various ways the Baroque spirit. The main square, for instance, represents the stage where people's life takes place, the point where one departs to go to work, the place of social activity and playing games and the site of civic and religious festivals. The plaza may also be viewed as a symbol of the "Theater of the World", where the deity, as the privileged spectator, contemplates the scene. This is characteristic of the Baroque idea of urban uses (Gutiérrez da Costa and Gutiérrez Viñuales 1995, 338).

Furthermore, religious constructions, especially the imposing churches which are thought to sacralise the space surrounding them, many of which have their altarpiece façade and open balcony-chapel, form the backdrop for an imaginary living performance. The same underlying considerations could be observed in the attempt to integrate certain green spaces into the urban plan, such as the kitchen garden and the orchard built behind the religious nucleus, in addition to the jungle which envelopes the whole urban plan. All these aspects are concrete expressions of the domain and control of the landscape so typical of the Baroque.

**The orientation of the missions**

Although it may be tempting to establish a close relationship between the missionary urbanism of Chiquitos and the ideas of Francesc Eiximenis, the Franciscan pioneer of humanistic Spanish urbanism, there are no records that endorse this connection. The work *Dotzé del Crestiá* (12$^{th}$ [Book] of the Christian), published in Valencian in 1384 by Eiximenis, offers precise instructions on how to design a city according to linear gridiron urban planning principles. In his instructions, Eiximenis proposed that the city-plan should take the form of a square since this is the most beautiful and ordered of all possible forms and enables the division of the city into a multitude of squared blocks (Vila Beltrán de Heredia 1987, 383). The 1573 Ordinances, mentioned above, reflect many of Eiximenis' guidelines, especially with regard to what constitutes favorable conditions at a site, including geographic aspects and particular orientation preferences; for instance, sites were to be cleared towards the south in order to take advantage of fresh southern winds, and planners were told to consider the availability of material resources – like the presence of water – and the security of the territory (Suárez Salas 1995, 410).

As already discussed, the urban layout of the Chiquitan missions departs from the way that was suggested in the Ordinances (and, also, from the grid-plan proposed by Eiximenis). It is also worth noting that while we do have "living" evidence of the urban planning of the missions today, there is no clear record that demonstrates that this planning respected any defined rule to orient towns or even their churches.

From Figure 4 we can see that the church of the mission was built with its axis parallel to the main entrance avenue to the village, the one that started from the Bethany chapel (*capilla Betania*, on the lower right of the figure), which is a small open oratory. However, this axis was not aligned with that avenue. It is a novelty in the missionary urbanism of Chiquitos that the axis of the main square and that of the village coincide with the axis of the College yard and this marks an important departure from the typical urbanism of the missions in the Province of Paraquaria, where it was customary that the church's axis coincided with that of the square (Roth 1995, 488; Giménez Benítez *et al.* 2018). The straight continuation of this avenue passed through the centre of the square (where we see the central stone cross surrounded by palm trees) and went on to the first courtyard (the main one, or the Jesuit courtyard) which bordered the church and where a sundial was placed. Indian houses covered three sides of the main square, while the fourth side was reserved for the religious nucleus. It must be emphasised though, that there are no records or chronicles written by the founding fathers themselves that indicate the significance of the spatial orientation of this great structure – that is, the village – either with regards to the surrounding landscape or to any other reference.

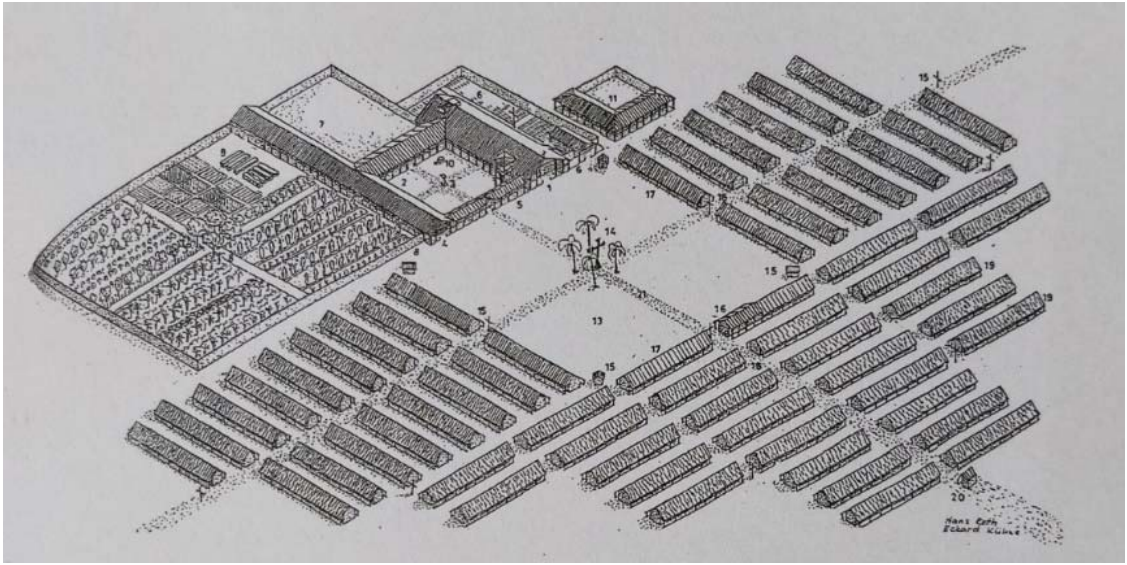

Figure 4. Ideal urban plan of a missionary village of Chiquitos, inspired by Concepción. Drawing by Hans Roth and Eckart Kühne, adapted from Roth, *Plano urbano ideal de un pueblo misional de Chiquitos, inspirado en Concepción*, 1995. In "Urbanismo y arquitectura en Chiquitos desde los testimonios materiales", in *Las Misiones Jesuíticas de Chiquitos*, edited by P. Querejazu (La Paz: Fundación Banco Hipotecario Nacional (BHN), 1995), p. 497. (Image courtesy of P. Querejazu Leyton).

**The Jesuit churches in Chiquitos**

The churches in this region are large simple pavilions with three naves and a wide, sloping gable roof. This covering was traditionally made with a wooden framework on which fired ceramic tiles were placed, seated with mud on a reed base. Typically, they are supported by 16 massive trunk columns obtained from trees of the rainforest, and around which helical spirals are carved to form Solomonic columns. These impressive columns are called *horcones* and can measure over 10 meters in length and weigh several tons. They constitute the backbone of the building structure.

According to Father José Cardiel, who left records for the churches of the Guaraní reductions, but which are equally valid for those of Chiquitos, "all these buildings are made in a different way than in Europe, because first the roof is made and then the walls" (Furlong 1953). This, of course, was not capricious but responded to the need to preserve the exterior walls, which, being made of adobe (clay mixed with straw), generally lack the needed resistance to torrential rains that hit these regions for long periods known as "times of waters".

The churches that we see today are very wide, and we know that – with some variations – the religious services in the Jesuit era usually hosted several thousand indigenous people, which was approximately the maximum number of reduced Indians in each missionary village. In this period, the churches were profusely embellished with various decorative elements which incorporated the architecture itself and other ornamental components (Martini 1977, 23). In the extension of the main nave there is a retired presbytery with several arches built of bricks and whose side walls are in line with the columns of the church. From there and through two vaulted openings, it links with the sacristies on both sides of the church. Outside of the building are roofed galleries and, in front of the temple façade, we find a spacious atrium with isolated columns. All this is clearly depicted in Figure 5, which shows the church of Nuestra Señora de la Inmaculada Concepción. The image on top conveys the size of the church, with the Sun-shaped oculus on the frontispiece, the large atrium and the lateral gallery. Also, the bell tower can be seen to the left of the construction. The bottom image shows the presbytery and some of the Solomonic-style columns that support the building.

**Methodology**

In our fieldwork we have measured the precise spatial orientation of these churches with the intention of verifying the possible existence of some definite pattern in the whole set. This specific analysis involves churches that belong exclusively to the Jesuit Order and which are located in a limited and very difficult to access territory (the tropical virgin forest). We think that the choice of locations with quite challenging geographical conditions is a unique characteristic of Chiquitan churches, a factor that could hardly be compared to Jesuit religious structures in Europe, North America or any other province that came under the influence of the Order. This study, therefore, provides us with the opportunity to verify whether the typical orientations found in Europe were rigidly transferred to the studied region or whether the adaptation of Christian churches to these new lands eventually followed other principles.

The work consisted of measuring the astronomical orientation (azimuth and angular height of the horizon) and the geographical location (latitude and longitude) at nine of the ten church sites. As noted above, the mission of Santo Corazón, which was moved to its current location after the Order's estrangement, could not be visited on our field trip and therefore its church was measured using satellite maps. We performed these measurements using a Suunto 360PC/360R tandem, which consists of a precision compass and clinometer, and we analysed the surroundings (landscape) of each of the buildings. The presented data is the result of several on-site measurements with a single instrument, taking the axes of the churches, from the back of the buildings towards the altars, as our main guide. We estimate measurement errors to be around 1/2°. In addition, geolocation technology (GPS Garmin eTrex 30) was used to obtain the precise position of each site.

The measured and calculated data for each of the churches studied are presented in Table 1, where we have ordered them by increasing azimuth (with values expressed in decimal degrees). For each church, its latitude and longitude (L and l), the astronomical azimuth (a), and name (which reflects the patron saint to which it is dedicated) are listed. Also included are the approximate dates of the documented first foundation and last transfer of the mission (dates next to the church names), and the most likely start date for the relevant church's construction placed in square brackets. Note that the original church of San Ignacio de Velasco was demolished in 1948 and rebuilt. Moreover, the current mission and church of Santo Corazón de Jesús are posterior to the expulsion of the Jesuits.

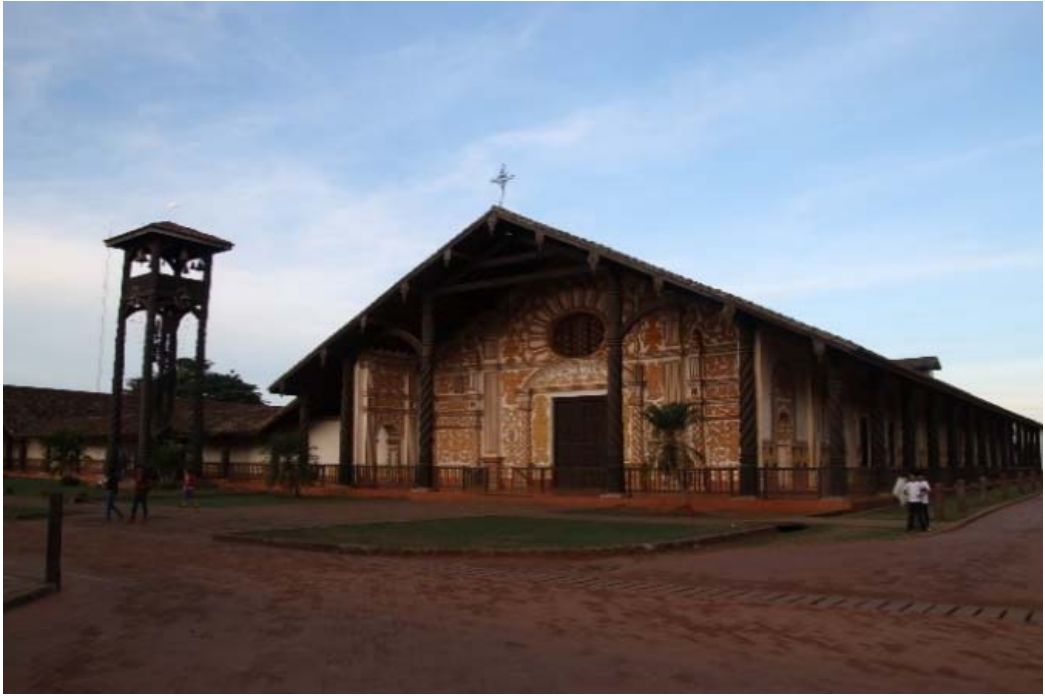

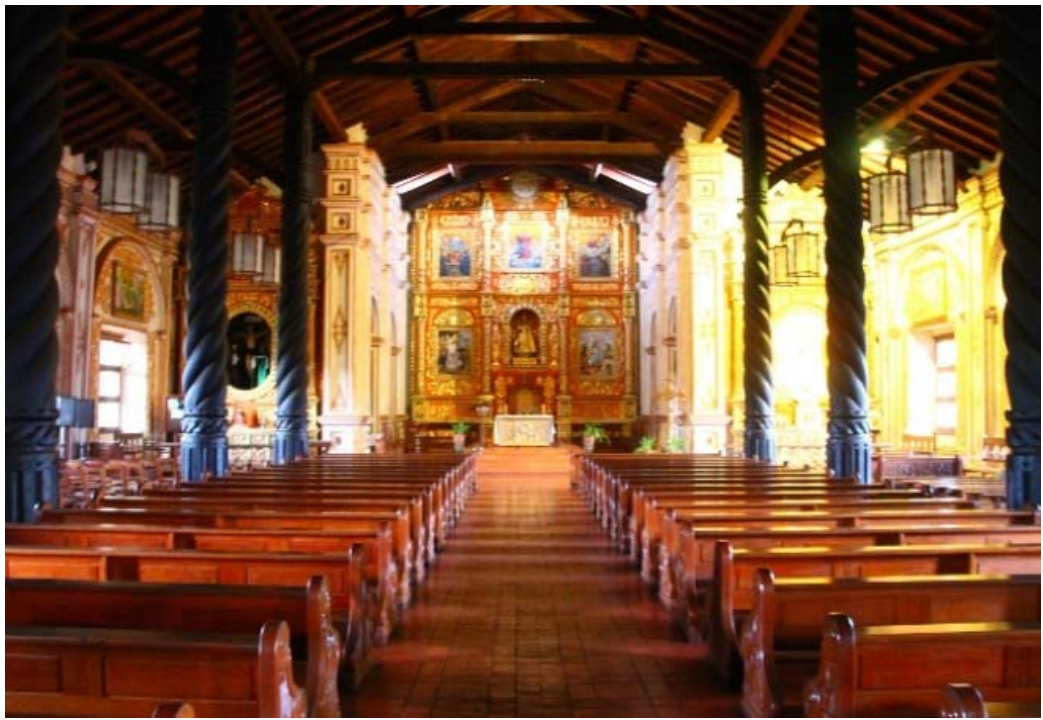

Figure 5. The church of Nuestra Señora de la Inmaculada Concepción. Photographs by the author.

The astronomical azimuth is calculated through the measured orientation of the axis of the construction towards the altar and then corrected for magnetic declination (Natural Resources Canada 2015). The magnetic declination values for the different church sites were between 12°52' and 14°39' W on the dates the measurements were made. The penultimate column includes the angular height of the point of the horizon (h), towards which the altar of the church is facing, corrected for atmospheric refraction (Schaefer 1993, 314).

In Table I, data for both the azimuth and the angular height of the horizon are rounded to 0.5º approximation. When the horizon was blocked (signalled as B in the penultimate column), we employed the digital elevation model based on the Shuttle Radar Topographic Mission (SRTM) available at HeyWhatsThat (Kosowsky 2017), which gives angular heights within a 0.5º approximation. Finally, in the last column, the calculated values of the resulting astronomical declination corresponding to the central point of the solar disc are reported.

| Name | L (º, S) | l (º, W) | a (º) | h (º) | δ (º) |
|---|---|---|---|---|---|
| San Rafael de Velasco 1696 1747 [1747] | 16.786602 | 60.674947 | 6.0 | +0.5 | +73.5 |
| Santa Ana de Velasco 1755 [c. 1760] | 16.583503 | 60.687550 | 10.0 | +0.5 | +71.2 |
| San Ignacio de Velasco 1748 [1748; new c. 1948] | 16.373513 | 60.960289 | 26.0 | 0.0 | +60.0 |
| San José de Chiquitos 1697 1706 [1725] | 17.845306 | 60.741700 | 90.5 | 0.0 | -0.6 |
| Ntra. Sra. de la Inmaculada Concepción 1708 1722 [1753] | 16.135558 | 62.023353 | 91.5 | -0.5 | -1.5 |
| San Juan Bautista (Taperas) 1699 1716 [1755] | 17.896234 | 60.376008 | 199.0 | B +1.5 | -66.3 |
| Santo Corazón de Jesús 1760 1788 [c. 1788] | 17.973958 | 58.807601 | 255.0 | B +2.5 | -15.2 |
| San Miguel de Velasco 1721 [1744] | 16.697787 | 60.968550 | 256.5 | 0.0 | -13.0 |
| San Francisco Javier 1691 1708 [1749] | 16.274553 | 62.505174 | 268.0 | 0.0 | -1.8 |
| Santiago de Chiquitos 1754 1764 [c. 1764] | 18.339631 | 59.598501 | 322.5 | +2.5 | +47.8 |

Table 1. Orientations of the Jesuit churches in the Province of Chiquitos.

**The orientation of churches and the landscape**

In Figure 6, we show the orientation diagram for the Jesuit churches of Chiquitos. The reported azimuths include the magnetic declination correction at each location. The diagonal lines on the eastern horizon of the graph indicate the extreme values of the corresponding azimuths for the Sun (azimuths of 65.6º and 114.8º – shown in solid-continuous lines, which are equivalent to the southern hemisphere winter and summer solstices respectively) and for the Moon (azimuths of 59.5º and 120.4º – shown in dotted lines, corresponding to the positions of the major lunistices or lunar standstills).

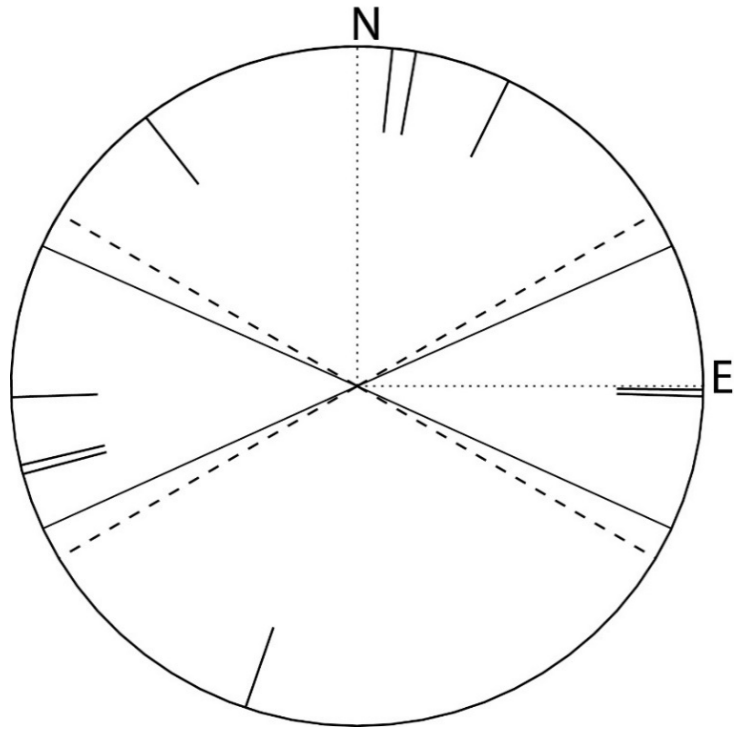

Figure 6. Orientation diagram of the Jesuit churches in the Province of Chiquitos, obtained from the combined azimuth data in Table 1.

Of the 10 measured orientations, four are directed towards the northern "quadrant" (that is, the northern corner with azimuthal orientations between 315º and 45º), and only one towards the southern quadrant (azimuth between 135º and 225º). The remaining five fall within the solar range, with three orientated towards the western horizon and two towards the eastern horizon.

It is notable that of these last five axis measurements, there are three that are, to a high degree, oriented towards the equinox (that is, with astronomical declination measures less than 2º). For example, San Francisco Javier which has an azimuth of 268º coincides (within the measurement uncertainties) with the geographical west. In the opposite direction, the axes of the churches of San José de Chiquitos and Nuestra Señora de la Inmaculada Concepción point towards the east with high precision, the first with an azimuth of 90.5º and the second with 91.5º. Although there are no textual sources that indicate the intentions of the founding priests in relation to the orientations of the Chiquitan missionary churches and despite the fact very few of them remain today for a statistical study to look for patterns, these results are still surprising.

On the other hand, the calculation of the astronomical declinations (listed in the last column of Table 1), does not present any novel or unanticipated findings. The heights of the horizon surrounding these churches generally have small "h" values, because most of the region consists of flat plains and soft hills. Furthermore, it is known that the Jesuits preferred sites that were elevated in relation to the local terrain when establishing new missions. This is demonstrated by the mission of San Miguel de Velasco (founded in 1721) which, according to Father Sánchez Labrador, who visited the mission before the expulsion of its Jesuit founders, had been located "in a good place, high and surrounded by forest" (Aguirre Acha 1933, 45). This naturally explains why the values of "h" are small and why a graph of declinations does not contribute more than is already apparent from the orientation diagram of Figure 6.

**Discussion**

Our results indicate that the orientations of the Jesuit churches in the Chiquitanía differ from the pattern found for the larger number of Guaraní churches (some 30 churches documented in the last period) that the Order erected in the Province of Paraguay. Previous research (Giménez Benítez *et al.* 2018; see also Ruiz Martínez-Cañavate 2017) has shown that most of the 27 ruins of the Jesuit churches that could be located in the northeast of Argentina and in the southern region of Paraguay and Brazil are orientated towards the southern quadrant (between 135º and 225º).

The orientation diagram in Figure 6 does not show any prominent concentration of azimuth distribution towards any specific quadrant, yet it does indicate that there are certain peculiarities that are worth noting. For example, only a tenth of the Guaraní churches analysed were oriented within the solar range, either westward or eastward (Giménez Benítez *et al.* 2018), while in the case of the churches of Chiquitos that proportion rises to half of the ten churches analysed.

Curiously, the church of Concepción (or rather, its ruins), located in the northeast of Argentina, is the only church among the 30 original Guaraní Jesuit churches that has an altar orientated towards the eastern quadrant and very close to the geographical east (Giménez Benítez *et al.* 2018). As can be seen in Table 1, the homonymous church in the Chiquitan town of Concepción is one of the only two orientated towards the east with great precision at an azimuth of 91.5º. Perhaps this is not a mere coincidence, since the missionary villages of Chiquitos began to emerge many decades after those of the Province of Paraquaria were established (c. 1632), and the relationship and exchange of documents between the Father Superiors of South American missions is well-known. This could have led the Swiss Jesuit architect Martin Schmid, who designed the church of Concepción in Chiquitos, to build the church there with an orientation similar to that of its Guaraní sister.

On the other hand, the organisation of the missions and the particular distribution of the different buildings within them, seems to have been a dynamic process, at least in the early years of each site. During the initial years of each mission, the churches were not the definitive ones; but once the priest in charge decided that the mission would remain in that place, then, they would complete and reinforce the earlier preliminary church and this is the "definitive" church that has survived up to now. It is also highly likely that when the time was ripe for reinforcing the definitive church, the village had already undergone some urban variations. One of the protagonists of the modern reconstruction and restoration of various churches, the Jesuit architect Hans Roth (Limpias Ortiz 2000), suggested that when Father Schmid orientated his churches "with the precision of a good Swiss mathematician and watchmaker" towards some cardinal point, he did so facing as a result the high cost incurred by the process of realigning the streets of the existing villages so that they followed the direction imposed by the axes of the new churches (Roth 1995, 484). Perhaps it is no coincidence, then, that the four churches built by Schmid are all oriented towards one of the four cardinal points: San Francisco Javier towards the west, Concepción towards the east – both with great precision – and San Rafael pointing just 6º east of the north. The axis of San Juan Bautista, in contrast, is almost 20º away from the south, but we must bear in mind that only the ruins remain of the original church (in particular, its bell tower, perhaps the oldest original non-restored construction existing today, which dates from 1755), and these are only partially excavated by archaeologists. Therefore, the measurements for this church are likely to contain errors.

The reduction of San Juan Bautista, located in Taperas, is also unique for other reasons. As we said, it remains as an archaeological ruin, but it is recognisable and is preserved in an extraordinary environmental site (Suárez Salas 1995, 423). The structure of the mission did not get altered in the republican period (early 19th century) after the expulsion of the Jesuits because, and for reasons that are unknown today, it was abandoned by its inhabitants in 1780, shortly after the expulsion of the Order. When D'Orbigny visited it during his journey in 1831 the church was still standing but in a very deteriorated condition. At the time, the explorer commented: "The tower was intact, but without a roof; in the church, which was very large, you could see the trunks of the trees grown next to columns of the same thickness" (D'Orbigny 1945, 1184). Although its imposing bell tower is still standing today, it is without a roof and flooded with weeds, and the church which was erected near it has disappeared almost without trace. Extensive future excavation work is needed to establish more reliable orientation measures of the axis of the church of San Juan Bautista than the preliminary measurements presented here.

The churches of San José de Chiquitos and San Miguel de Velasco are both orientated within the solar range and, as we pointed out, the former is equinoctial to a remarkably high precision (azimuth of 90.5º). San Ignacio de Velasco, on the other hand, is oriented towards the northern quadrant (azimuth of 26º), very far from the solar range. As is the case with the other churches, the reasons for this orientation were not documented in the extant chronicles of the time. Unfortunately, San Ignacio Church, which according to Roth (1995, 466) was "the most beautiful of its style in the interior of America", endured a harsher fate than the other churches in the region, for due to the lack of maintenance and prolonged abandonment over the decades, it had to be demolished in 1948. While Roth's opinion was subjective, textual evidence confirms that it was the largest, most elaborate and ornate church in the region. This could have rendered its restoration a difficult and arduous task. A short time later, on the same site of the demolished building, the present church which stands today was built.

The missions of Santiago, Santa Ana and Santo Corazón were founded during the last period of the Jesuit presence in Chiquitos, between 1750 and 1766, to accommodate the large numbers of indigenous people who agreed to be reduced during this period. It is important to point out that the mission of Santo Corazón had to be relocated after the 1767 expulsion, and so, the construction of the church was not supervised by members of the Order despite its apparent Jesuit-style architecture. Moreover, it should be noted that the churches of the first two of these three "late" missions have their axes pointing towards the northern quadrant – in the case of Santa Anta with an azimuth even lower than that of San Ignacio – and far away from the solar range. Although the reasons for these orientations cannot yet be discerned, it is likely that the orientation of these villages and churches was neglected due to the urgent need to shelter large numbers of indigenous peoples. The village of Santa Ana, in particular, shows a clear master axis that begins at the Bethany chapel, then passes through the stone cross of the main square and finally passes through the sundial and part of the College, but differs by a few degrees from the axis of its church (Roth 1995, 488; Suárez Salas 1995, 431). This can be verified both through archaeological maps of the mission and by satellite images, since Santa Ana preserves many of its historical structures and original distribution of spaces.

We saw in Figure 4 that the Bethany chapel was located at the opposite end of the religious nucleus and marked the beginning of the Chiquitan village via its alignment with the main avenue. In Chiquitos, the Palm Sunday processions started there and headed towards the main square. We know from the New Testament – Gospels of Matthew 21, Mark 11 and Luke 19 – that Bethany derives its name from a little town located at the foot of the Mount of Olives from where Jesus rode his donkey to make his triumphal entry into Jerusalem. The original biblical town, today called al-Eizariya (literally, place of Lazarus), is located towards the east of Jerusalem. So, if the Jesuits had wanted to emulate this biblical geography on a reduced scale in the missionary villages (and in their churches), they should have oriented all of them in a deliberate and similar pattern, with the altars of the churches facing west. Clearly, however, this is not the case.

We mention now a topic that we believe warrants a detailed future study, and which could also be related to the orientation of the mentioned churches. We refer to the illumination effects within the churches, that is, how natural light, or sunlight, was employed to highlight and accentuate particular features of the architecture and its interiors, such as the altar (see, for example, the recent study by Vilas-Estévez and González-García 2016, for the Cathedral of Santiago de Compostela). The churches of Chiquitos were characterised by low windows in the side walls, at a level corresponding roughly to the height of the people, and well protected by covered galleries of wide corridors; both distinct features rendered the Chiquito churches relatively dark. However, the inseparable relationship between Baroque architecture and natural light is well known and must not be dismissed in future investigations. A manifestation of the mentioned illumination effects could be witnessed in churches like San Miguel and San Rafael, where bright *lucernarios*, a kind of skylight located on the rooves, illuminate the presbyteries of the churches. Due to the northern orientation (azimuth of $6^0$) of San Rafael, the light of the tropical midday Sun does not enter through the side windows or through the main door of the church. Therefore, the presence of skylights allows the altar to be illuminated with natural light (Figure 7). The black and white photograph by Hans Erlt (bottom image of Figure 7) shows the evening beam of light as it passes through one of the skylights into the presbytery and illuminates the main altarpiece.

Describing San Rafael Church, Roth pointed out: "In summer, from November to February, a ray of the sun slowly crosses the main altarpiece of San Rafael, entering through the two skylights on the roof" (Roth 1995, 507). Despite the silence of historical Jesuit chronicles, it is very likely that the Father architects of these churches might have considered the diurnal journey of the Sun at different times of the year in the orientation, design and construction processes. This would have allowed them to take advantage of the sunlight to either illuminate the chancel or to create other notable effects.

In addition, the known veneration of the Chiquito Indians for the most prominent stars, the Moon and the Sun, is reflected in syncretistic sculptures and paintings which were once present on the altarpieces of the churches (Roth 1995, 507). Furthermore, there is evidence that the Sun was used to create a special atmosphere or climax during the liturgy. For example, in east-oriented churches such as Concepción and San José de Chiquitos, and especially at equinoctial times of the year, a marvellous effect was produced during the afternoon Mass when the setting Sun entered the church and bathed with its golden hue the polychromes of the main altarpiece and the figure of the patron saint. A similar effect was probably encountered in the church of San Francisco Javier (and at other times of the year also in San Miguel and in Santo Corazón) but during the morning service.

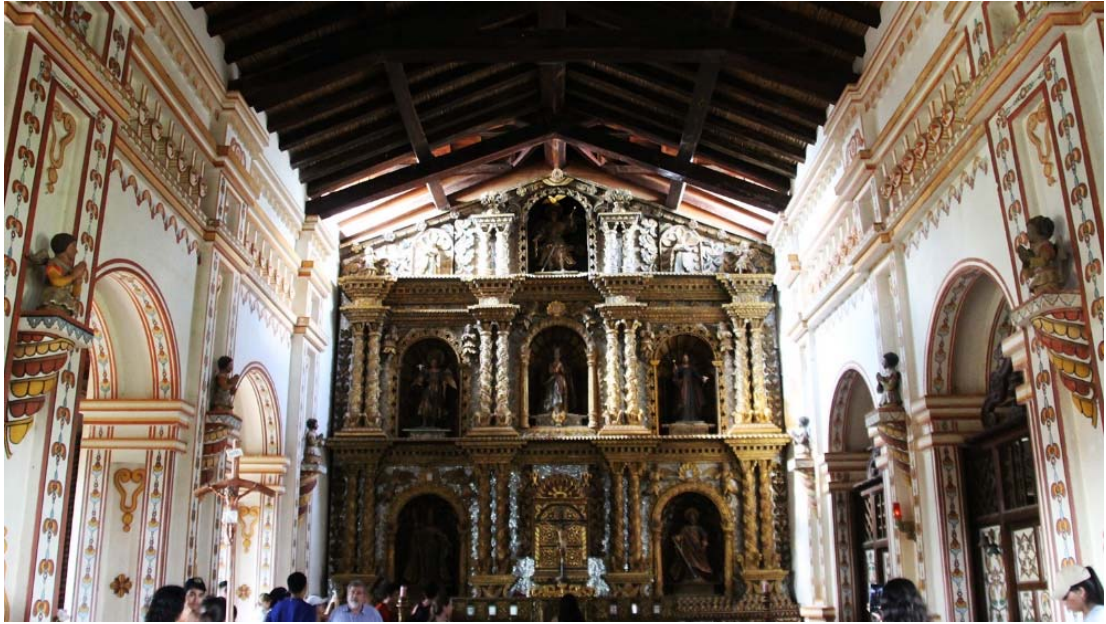

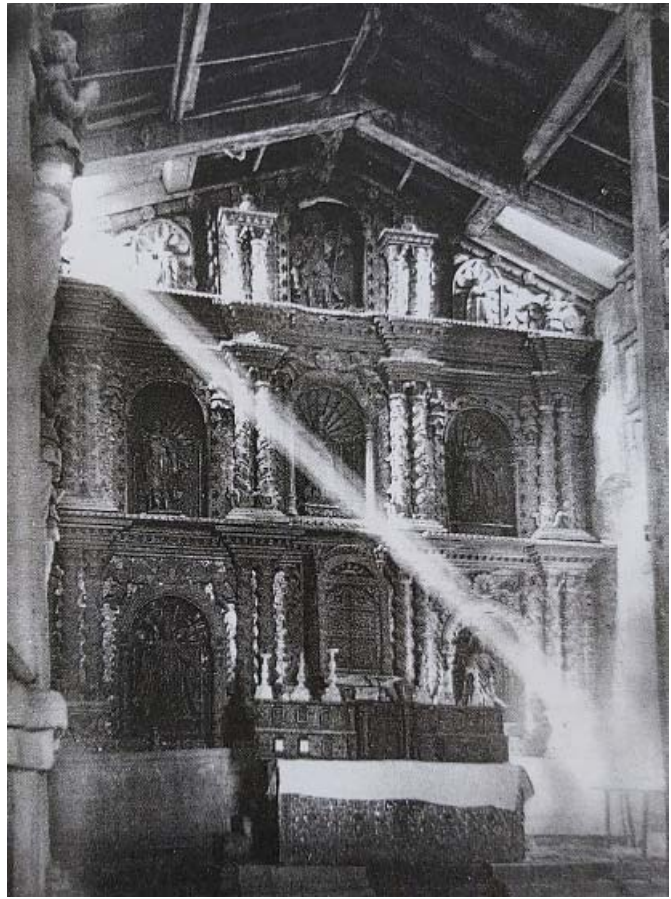

Figure 7. *Lucernarios* of San Rafael. Top: photograph by the author. Bottom: a photograph of 1951 by Hans Erlt, adapted from *Retablo mayor de San Rafael en cedro, tallado y dorado,* 1995. In "Archivo Fotográfico de Chiquitos", in *Las Misiones Jesuíticas de Chiquitos*, edited by P. Querejazu (La Paz: Fundación Banco Hipotecario Nacional (BHN), 1995), p. 111. (Photo courtesy of P. Querejazu Leyton).

The presence of an "oculus" – a high window shaped like the Sun – on the frontispieces of three of the ten mission churches (San Francisco Javier, Concepción and San Rafael, all built by Father Schmid), is a clear indication of the deliberate consideration of illumination effects during the construction of the churches (Figure 5, above). Decorative functions aside, these openings were undoubtedly designed to make the most of the entry of sunlight during religious services.

There are detailed studies of illumination effects in many churches and cathedrals around the globe (for example Heilbron 1999, 93, in the context of *meridiane*, and the aforementioned work by Vilas-Estévez and González-García 2016), yet only scattered comments about the same survive in the case of the churches in Chiquitos; an issue that undoubtedly necessitates further research.

**Conclusion**

In this study, we measured and analysed the orientations of the whole group of surviving Jesuit churches in the Chiquitos missions of eastern Bolivia. Our main aim was to look for a possible pattern of orientation of these iconic buildings and to see whether it could be related to the varying positions of sunrises and sunsets throughout the year. To provide the study with adequate contextualisation, we commenced with a detailed cultural and historical study of the Jesuit missions in the region and of the characteristics of the villages where the churches are located.

Our results show that despite the small number of Jesuit religious constructions still standing in Chiquitos, it is possible to draw some interesting conclusions. Remarkably, we found that unlike what was indicated by previous studies for the churches of the Province of Paraquaria, where the vast majority of those churches had their axes pointing towards the southern part of the horizon (Ruiz Martínez-Cañavate 2017, 168; Giménez Benítez *et al.* 2018), half of the studied Chiquitos churches are oriented within the solar range.

Furthermore, three of these historical constructions were found to have a very precise equinoctial orientation with an error of less than 2°. These include the churches of Nuestra Señora de la Inmaculada Concepción and San José de Chiquitos, whose altars point towards the east, and San Francisco Javier, where the altar faces westward. As mentioned in the discussion, there remain no written records on the intentions of the founding priests that could explain the found results. It must be further noted that, as we also discussed, two of these equinoctial churches were constructed by Father Martin Schmid, who tended to orient the churches towards the cardinal points. Therefore, this could provide a prosaic answer, at least for a few churches.

We leave for a future investigation the possible influence of the particular custom established by the Order for the missions, referred to in textual sources as the "modo nostro" ("our way"). The particular mode of construction paid special attention to community needs and to the practical functionality of spaces for liturgical purposes, and reflected the Order's flexibility and adaptability to the peculiarities of each territory. Hence, it is possible that there could have been a fusion or adaptation of the instructions dictated from Rome – in terms of the design of the missions and even the orientation of the temples – with the usual customs typical of each place. One should pay special attention to the possibility that the layout of the churches was not completely independent of the urban grid of the villages, and particularly of the location of the main squares of the missions (although Roth suggested that Schmid's churches may have involved adjusting the streets to accommodate the church). These elements, characteristic of Jesuit thinking, might also explain the diversity of orientations that were measured, with churches that direct their altars in the traditional "canonical" way in the minority.

We have also briefly commented in this paper on the possible relevance of natural illumination effects on certain internal church structures and elements, highly sought after in Baroque architecture, that could have influenced the orientation preferences of the builders of these churches. Thus is another important element worthy of further research, since natural illumination could explain the orientation of some individual constructions of the Jesuit Order in the region.

Finally, based on the analysis of the data presented above, it could be concluded that the orientation of the Jesuit churches of Chiquitos (and of the missions themselves) most likely does not follow a well-defined prescription or pattern, since it appears to us that the churches had to be adapted to each individual site and landscape, which posed a number of challenges. This notwithstanding, from the measurements and subsequent analysis it is possible to discern certain notable characteristics in this group of churches that it has been worth highlighting.

**Acknowledgements**

The author wishes to thank Professor Víctor Hugo Limpias Ortiz, from UPSC (Santa Cruz de la Sierra), for his kind invitation to work in Chiquitos, and Drs Limpias Ortiz and Pedro Querejazu for the many discussions on Chiquitan culture during the visits to the missions. The author also thanks José X. Martini, Virgilio Suárez Salas and Eckart Kühne for their various suggestions and recommendations, and Sixto Giménez Benítez for discussions about the Guaraní Jesuit churches and for allowing the author to read a preliminary version of his work prior to its final publication. This work has been partially supported by CONICET and by the University of Buenos Aires.